\documentstyle[prd,aps,epsfig,eqsecnum,amssymb,amsmath,floats]{revtex}
\begin{document}
\title{Search for solar antineutrinos and constraints on the neutrino background 
from PBHs}
\author{
E. V. Bugaev and K. V. Konishchev
}
\address{
Institute for Nuclear Research, Russian Academy of Sciences, Moscow 117312,
Russia
}
\date{
\today
}
\maketitle
\begin{abstract}

We obtain the constraints on the diffuse neutrino background from primordial 
black hole evaporations using the recent data of KamLAND experiment setting 
an upper limit on the antineutrino flux from the Sun.

\end{abstract}

\section{Initial PBH mass spectrum}

According to Press-Schechter model, the fraction of mass in the universe 
contained in gravitationally bounded object more massive than $M$ is \cite{PS}
\begin{equation}
\label{PS}
\frac{1}{\rho_i}\int\limits_M^{\infty} M\frac{dn}{dM}dM = P (\delta_R>\delta_c, M, t_i).
\end{equation}
Here, $\frac{dn }{dM} dM$ is the number density of objects of mass in $(M,M+dM)$,
and the integral on the left-hand side gives the total mass density in overdense
regions more massive than $M$. $P(\delta_R, M, t_i)$ is the probability that a randomly
chosen point in the universe has density contrast $\delta_R> \delta_c$ at time $t_i$, 
and $M$ is the smoothing scale of $\delta_R$ ($R$ 
is the radius of the spherical overdense region).
The value $\delta_c$ is some critical value of the density contrast which corresponds to the
beginning of the non-linear regime. In a case of  a gaussian field of density fluctuations
this probability is 
\begin{equation}
\label{__2}
P=\frac{1}{2}erfc\left(\frac{\delta_c}{\sigma_R\sqrt{2}}\right),
\end{equation}
where $\sigma_R^2=\sigma_R^2(M,t)=<\delta_R^2>$ is the variance of the smoothed 
density. The mass distribution of gravitationally bounded objects is \cite{PS}
\begin{equation}
\label{__3}
\frac{dn }{dM}=\frac{2\rho_i}{M}\frac{\partial P}{\partial M}.
\end{equation}
The variance $\sigma_R^2$ is given by the relation \cite{ARLDHL}
\begin{equation}
\label{__4}
\sigma_R^2(M,t)=\int\limits_0^{\infty}\frac{k^4}{(aH)^4}\delta_H^2(k,t)W^2(kR)T^2(k)\frac{dk}{k},
\end{equation}
where $W(kR)$ is the smoothing window function, $\delta_H (k,t)$ is the horizon crossing 
amplitude,  $T(k)$ is the transfer function.

For the calculation of the mass spectrum of PBHs one must use the connection between the 
PBH mass $M_{BH}$ and characteristics of the overdense region. In the model of the 
near-critical collapse proposed in ref.\cite{16} one has the relation: 
\begin{equation}
\label{__5}
M_{BH}= M_{i}^{1/3}M^{2/3}k(\delta_H^R - \delta_c)^{\gamma_k},
\end{equation}
where $M_i$ is the horizon mass at  time $t_i$ when the density fluctuations develop,
and $\delta_H^R$ is the smoothed density contrast at the moment when the fluctuation crosses 
horizon,
\begin{equation}
\label{__6}
\delta_H^R=(M/M_i)^{2/3}\delta_R.
\end{equation}

In the approach of ref.\cite{PH} all PBHs have masses roughly equal to the horizon
mass at a moment of the formation, independently of the
perturbation size. The approximate connection between the PBH mass 
and the horizon mass in such a model is 
very simple:
\begin{equation}
\label{__7}
M_{BH}=\gamma^{1/2}M_i^{1/3}M^{2/3}.
\end{equation}
where $\gamma$ is the ratio of the pressure to energy density ($\gamma=1/3$ in the 
radiation dominated era).

Using these formulas one obtains the final expression  for the PBH mass spectrum
\cite{1}

\begin{eqnarray}
\label{1_20}
n_{BH}(M_{BH})=\sqrt{\frac{2}{\pi}} {\rho_i} \int\frac{1}{M\sigma_H  (M_h)}
\left|\left\{
        \frac{2}{3M}-
	\left[
	    \frac{n'}{2}ln \frac{k_{fl}}{k_0}+
	    \frac{n(k_{fl})-1}{2}\frac{1}{k_{fl}}
	\right] 
\cdot\frac{\partial k_{fl}}{\partial M}
        \right\}\right.       \times\nonumber\\
\\
\left.\left(\frac{(\delta_H^R)^2}{\sigma_H^2 (M_h)}-1\right)\right|
e^{-\frac{(\delta_H^R)^2}{2 \sigma_H^2 (M_h) }}\frac{d\delta_H^R}{d f(M,\delta_H^R)/dM}\;.
\nonumber
\end{eqnarray}
Here we use the general notation for the connection between $M_{BH}$, $M$ and $\delta_H^R$,
\begin{equation}
\label{__8}
M_{BH}=f(M,\delta_H^R).
\end{equation}
Particular cases of this connection are given by Eqs.(\ref{__5},\ref{__7}).

Substituting the expression for the window function,
\begin{equation}
\label{__9}
W(kR)= 3\left(\frac{\sin{kR}}{(kR)^3} -\frac{\cos{kR}}{(kR)^2} \right),
\end{equation}
and for the transfer function \cite{_1},
\begin{equation}
\label{__10}
T(k)\approx W(\frac{kR}{\sqrt{3}}),
\end{equation}
and the parametrization 
\begin{equation}
\label{__12}
\delta_H(k)=\delta_H(k_0)\left(\frac{k}{k_0}\right)^{\frac{n(k)-1}{2}}
\end{equation}
for the horizon crossing amplitude into Eq.(\ref{__4}) for the variance $\sigma_R^2$
one obtains, approximately,
\begin{equation}
\label{__13}
\sigma_R^2(M)=\left(\frac{M}{M_i} \right)^{-4/3}\sigma_H^2(M)\approx 
\frac{k_{fl}^4}{(aH)^4}C^2 \delta_H^2(k_{fl}).
\end{equation}
The exact value of the coefficient $C^2$  depends on the spectral index (for $n=1.3$ 
$C^2=6.82$). 
In Eq.(\ref{__13}) $k_{fl}$ is the comoving wave number, characterizing the perturbed region, 
$k_{fl}=\frac{1}{R}$, and $M_h$ is the fluctuation mass at the moment when the 
perturbed region enters the horizon at
the radiation dominated era. The value of $M_h$ is simply connected with $k_{fl}$:
\begin{equation}
\label{__15}
k_{fl}=a_{eq}H_{eq}\left(\frac{M_{eq}}{M_h}\right)^{1/2}.
\end{equation}
At the same time, an initial value of the fluctuation mass, $M$, is proportional to $R^3$,
so $M_h\sim M^{2/3}$. The exact connection is
\begin{equation}
\label{__16}
M_h=M_i^{1/3}M^{2/3}.
\end{equation}
The denominator $(aH)^4$ in Eq.(\ref{__13}) must be taken at $t=t_i$, and one can easily see that
\begin{equation} 
\label{__17}
\frac{k_{fl}^2}{(a_iH_i)^2}=\left(\frac{M}{M_i}\right)^{-2/3}.
\end{equation}
At last, in the spectrum formula (\ref{1_20}) the following notation is used
\begin{equation}
\label{__18}
n'=\left.\frac{d n(k)}{dk}\right|_{k=k_{fl}}.
\end{equation}
The limits of integration in Eq.(\ref{1_20}) are determined using the conditions
\begin{equation}
\label{1_19}
\delta^R_H{}_{min}=\delta_c\;\;\;\;\;\;,\;\;\;\;\; M_{min}=M_{h}^{min}=M_i\;\;\;.
\end{equation}

We have, for calculations of $n_{BH}(M_{BH})$
two free parameters: the spectral index $n$, giving the perturbation amplitude 
through the normalization on COBE data on large scales, and $t_i$, 
the moment of time just after reheating, from
which the process of PBH formation started. The value of $t_i$ is 
connected, in our approach, with a value of the
reheating temperature,
\begin{equation}
\label{7_}
t_i=0.301 g_*^{-1/2}\frac{M_{pl}}{T_{RH}^2}
\end{equation}
($g_*\sim 100$ is a number of degrees of freedom in the early Universe).
The  initial horizon mass $M_i$ is expressed through~$t_i$:
\begin{equation}
\label{7a_}
M_i\cong \frac{4}{3}\pi  t_i^3\rho_i.
\end{equation}

In the case of Page-Hawking \cite{PH} collapse the minimum mass in the 
PBH mass spectrum is given by the simple formula
\begin{equation}
\label{__20}
M_{BH}^{min}=\gamma^{1/2}M_i
\end{equation}
which follows from Eq.(\ref{__7}). Correspondingly, the minimum value of 
PBH mass is determined entirely by the reheating temperature. 
In the case of  the near-critical collapse there is no minimum value
of the PBH mass in the spectrum because $\delta^R_H$ can be almost equal
to $\delta_c$ (see Eq.(\ref{__5})). In this case the PBH mass spectrum 
has a broad maximum at  $M_{BH}\sim M_i$.

\section{Neutrino diffuse background from PBHs}

Evolution of a PBH mass spectrum due to the evaporation leads to the
approximate expression for this spectrum at any moment of time:
\begin{equation}
\label{8}
n_{BH}(m,t)=\frac{m^2}{(3\alpha t + m^3)^{2/3}} n_{BH} \left((3\alpha t +
m^3)^{1/3}\right),
\end{equation}
where $\alpha$ accounts for the degrees of freedom of evaporated particles and,
strictly
speaking, is a function of a running value of the PBH mass $m$. In all our
numerical
calculations we use the approximation
\begin{equation}
\label{9}
\alpha=const=\alpha (M_{BH}^{max}),
\end{equation}
where $M_{BH}^{max}$ is the value of $M_{BH}$ in the initial mass spectrum
corresponding to a maximum of this spectrum or, in the case of Page-Hawking collapse,
to a minimum value of PBH mass in the spectrum. Special study shows that errors
connected with such an approximation are rather small.
The detailed calculations of the function $\alpha(m)$ were carried out in the 
works \cite{22}. Here we use the simplified approach in which $\alpha(m)$ is
represented by a series of step functions \cite{11}

The expression for a spectrum of the background  radiation is \cite{11}
\begin{eqnarray}
\label{10}
S(E)=\frac{c}{4\pi}\int dt \frac{a_0}{a}\left(\frac{a_i}{a_0}\right)^{3}
\int dm \frac{m^2}{(3\alpha t + m^3)^{2/3}} n_{BH} \left[(3\alpha t +
m^3)^{1/3}\right]
\cdot f(E\cdot (1+z),m)e^{-\tau(E,z)}\nonumber\\
\\
\equiv \int F(E,z)d \log_{10} (z+1).\nonumber
\end{eqnarray}

In this formula
$a_i$, $a$ and $a_0$ are cosmic scale factors at $t_i$, $t$ and at   present
time,
respectively, and
$f(E,m)$ is a total instantaneous spectrum of the background radiation
(neutrinos or photons) from the black hole evaporation. It includes the pure
Hawking
term and contributions from fragmentations of evaporated quarks and from decays
of
pions and muons (see \cite{11} and  earlier papers \cite{22,23,24} for details).

In the last line of Eq.(\ref{10}) we  changed the variable $t$ on $z$ using
the flat model with $\Omega_{\Lambda}=0$
 for which
\begin{eqnarray}
\label{11}
\frac{dt}{dz}=-\frac{1}{H_0 (1+z)}\left(\Omega_m (z+1)^{3}+\Omega_r
(z+1)^{4}\right)
^{-1/2},\nonumber\\
\\
\Omega_r = (2.4\cdot 10^{4} h^2)^{-1} \;\;\;,\;\;\; h=0.67.\nonumber
\end{eqnarray}

\section{Results}
\label{sec:Dis_and_Con}

In some theoretical schemes (e.g., in the model of spin-flavor precession in a magnetic field)
the Sun can emit a large flux of antineutrinos. The LSD experiment
\cite{26} sets the upper limit on this flux, $\Phi_{\tilde \nu}/\Phi_{\nu}\le 1.7$~\%.
The corresponding constraints on diffuse neutrino background from primordial black holes
were obtained in the work of authors \cite{11}.

Recently, the KamLAND experiment obtained a new limit on electron antineutrino flux 
from the Sun, according to which this flux is less than $2.8\times 10^{-4}$
of the SSM ${}^8B$~$\nu_e$ flux. The neutrino detection was carried out
in both experiments using the reaction 
\begin{equation}
\label{27.1}
\tilde \nu_e + p \to n + e^{+}.
\end{equation}
No candidates were found in \cite{3} for an expected background 
of $1.1\pm 0.4$ events. We calculate the limit on the antineutrino flux
using the formula 
\begin{equation}
\label{27.2}
\Phi_{\tilde \nu_e}=\frac{1.1}{\bar\sigma\cdot\Delta E\cdot T\cdot N_p\cdot 4\pi},
\end{equation}
where $\sigma$ is the cross-section of the reaction (\ref{27.1}) averaged over
neutrino background spectrum from PBHs, $\Delta E$ is the neutrino energy range
measured in the experiment ($8.3-14.8$~MeV), $T=1.6\cdot 10^{6}$~s is the livetime,
and $N_p=4.6\cdot 10^{31} $ is the number of target protons in the fiducial volume.

Using the results of calculations of electron neutrino background spectra from
PBHs (the examples of such spectra are given on Fig.1 for the case of Page-Hawking
collapse) and normalization of the perturbation amplitude $\delta_H (k)$ on
COBE data (for details see \cite{11}), we obtain the constraints on the spectral
index $n$ shown on Fig.2.

\begin{figure}[!t]
\epsfig{file=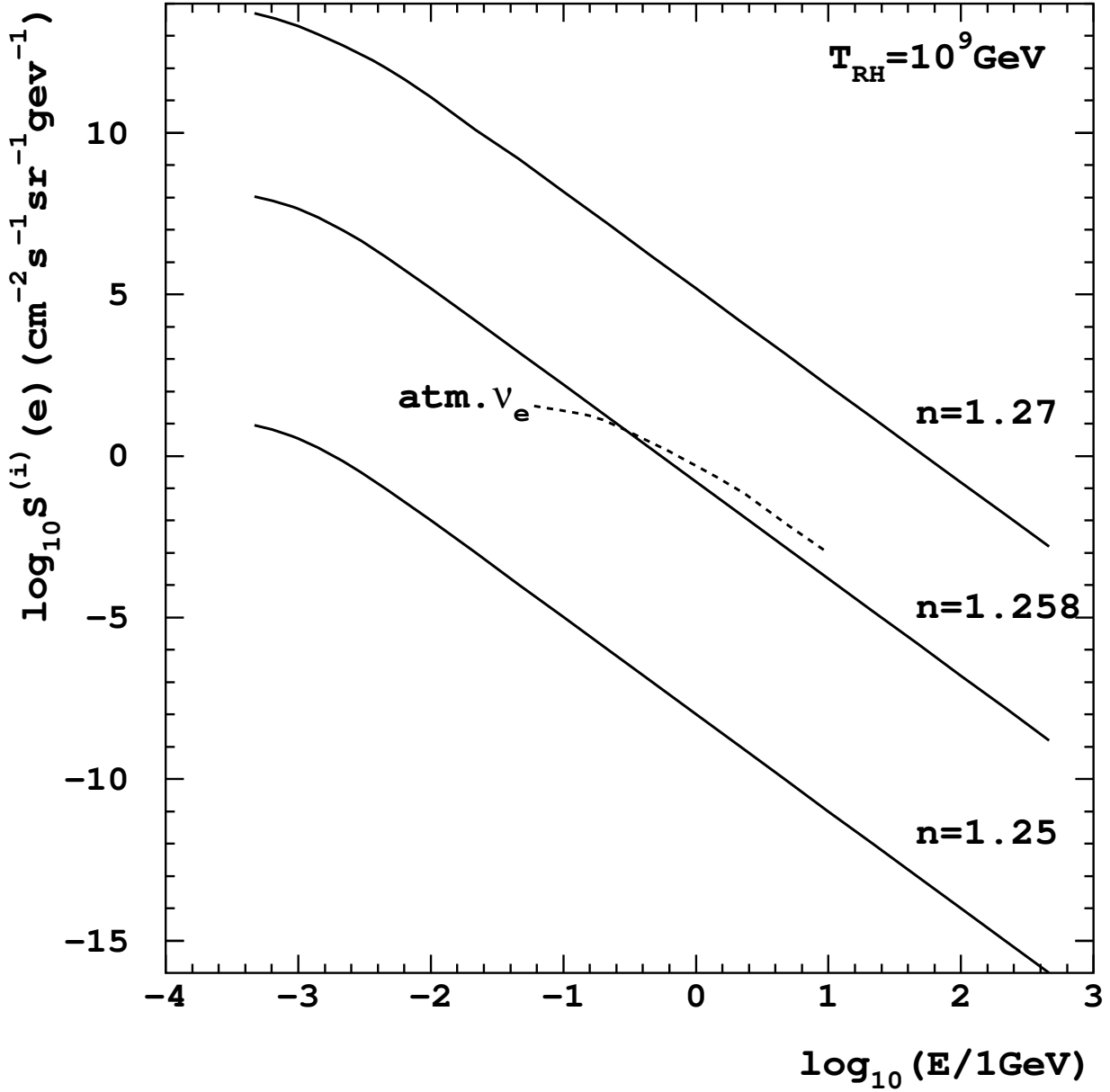,width=\columnwidth}
\caption{Electron neutrino background spectra from PBHs, calculating for several values
of the spectral index. Dashed curve shows the theoretical atmospheric neutrino spectrum at
Kamiokande site [17] (averaged over all direction)}
\label{fig:fig1}
\end{figure}

\begin{figure}[!t]
\epsfig{file=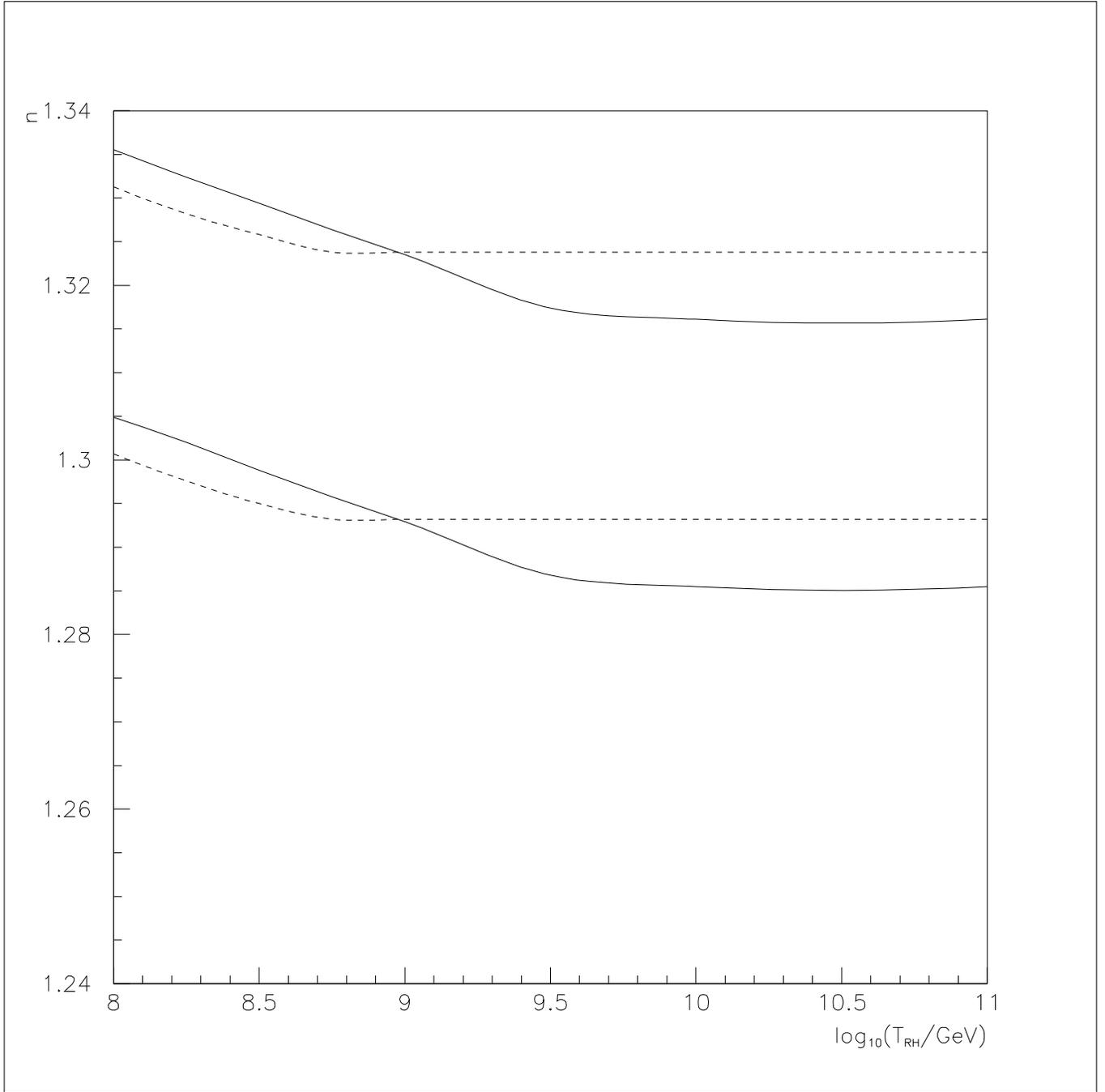,width=\columnwidth}
\caption{Constraints on spectral index from KamLAND experiment data (solid
lines) in comparison with constraints from data on extragalactic gamma ray background 
(dotted lines). Upper lines correspond to
critical collapse scenario of the PBH formation, lower ones - to standard case.}
\label{fig:fig2}
\end{figure}

\end{document}